\input epsf
\documentstyle[prd,aps]{revtex}
\begin{document}
\draft

\title{The Evolution of Distorted Rotating Black Holes III: Initial Data}

\author{Steven R. Brandt${}^{(1,2)}$
 and Edward Seidel${}^{(1,2)}$}
\address{${}^{(1)}$ National Center for Supercomputing Applications \\
605 E. Springfield Ave., Champaign, Il. 61820}

\address{${}^{(2)}$ Department of Physics
University of Illinois at Urbana-Champaign}

\date{\today}

\twocolumn[
\maketitle
\begin{abstract}
\widetext
In this paper we study a new family of black hole initial data sets
corresponding to distorted ``Kerr'' black holes with moderate rotation
parameters, and distorted Schwarzschild black holes with even- and
odd-parity radiation. These data sets build on the earlier rotating
black holes of Bowen and York and the distorted Brill wave plus black
hole data sets.  We describe the construction of this large family of
rotating black holes.  We present a systematic study of important
properties of these data sets, such as the size and shape of their
apparent horizons, and the maximum amount of radiation that can leave
the system during evolution.  These data sets should be a very useful
starting point for studying the evolution of highly dynamical black
holes and can easily be extended to 3D.
\end{abstract}
]


\narrowtext

\section{Introduction}
\label{intro}

There is increasing interest in the study of black holes from both
astrophysical and theoretical points of view.  Coalescing black hole
binaries are considered an important source of gravitational waves for
LIGO~\cite{Abramovici92}, VIRGO~\cite{Thorne94a} and other
gravitational wave observatories.  At the same time, numerical studies
of black holes are advancing to the point where highly distorted,
axisymmetric black holes~\cite{Abrahams92a,Anninos93c} and black hole
collisions~\cite{Anninos93b,Anninos94b} can be simulated, extracting
important physics such as the gravitational wave forms emitted and the
dynamics and properties of the event and apparent
horizons~\cite{Anninos94f,Matzner95a}.  Recent progress has also been
made in 3D~\cite{Anninos94c} black hole spacetimes.  These are some of
the reasons it is important to develop a series of initial data sets
to evolve.

However, black hole initial data sets are interesting in their own
right, apart from evolutions. They are potentially useful for studying
problems related to cosmic censorship, such as the Penrose
inequality~\cite{Bernstein93a}. Furthermore, it is possible to study
greater numbers of data sets with higher resolution than is practical
to evolve numerically.  This can lead to important observations of
properties of apparent horizons and can help to identify particularly
interesting spacetimes for evolution studies.  Initial data sets of
the type described in this paper can be used to analyze the most
extreme types of distortion and may have bearing on the hoop
conjecture~\cite{Thorne72a,MisnerHoopRef}.  The study of an apparent
horizon in an initial data set may provide valuable insight into which
spacetimes might have the most distorted event horizons.

Black hole initial data sets, based on the Einstein-Rosen bridge
construction~\cite{Einstein35}, have been developed over the last few
decades.  In the early 1960's a series of time symmetric, conformally
flat wormhole data sets corresponding to two black holes were
developed by Misner~\cite{Misner60} and Brill and
Lindquist~\cite{Brill63}.  At about the same time Brill developed an
initial data set for gravitational wave spacetimes that involved
specifying a free function in the conformal part of the 3--metric that
determined the distribution of the gravitational wave energy in the
spacetime~\cite{Brill59}.  These time symmetric initial data sets were
combined to produce the NCSA ``Brill wave plus black hole''
spacetimes, generalizing the single wormhole to a highly distorted
black hole spacetime~\cite{Bernstein93a,Bernstein94a}.  The evolution
of these data sets has been very useful in understanding the dynamics
of highly distorted black hole spacetimes, and also in preparing for
studies of colliding black holes, as colliding holes form a highly
distorted single hole immediately after the merging process.
Rotating, stationary black hole data sets, which are not
time-symmetric, were first discovered by Kerr~\cite{Kerr63} in 1963,
and then a nonstationary form was discovered by Bowen and
York~\cite{Bowen80}.  The conformally flat Bowen and York construction
for single black holes was then generalized to multiple conformally
flat holes with angular momentum by a number of authors (for example,
see~\cite{Cook90}) leading to recently computed 3D data sets for two
black holes with arbitrary spin and angular momentum~\cite{Cook94}.

In this paper we discuss a new family of distorted rotating black hole
initial data sets, some of which were evolved and studied in
Ref.~\cite{Brandt94b,Brandt94c,Brandt94d}. Although the constructions
we present here are axisymmetric, our approach is easily carried out
in 3D.  These data sets combine a number of the ideas discussed above,
generalizing the Bowen and York construction by including a ``Brill
wave'' so that highly distorted, rotating black holes with
gravitational radiation can be studied. In addition to these new data
sets to be described below, this family includes as special cases all
previous single black hole data sets discussed above, including the
Schwarzschild, Kerr, Bowen and York, and NCSA ``Brill wave plus black
hole'' spacetimes.

As we have shown in Ref.~\cite{Anninos94f,Brandt94c,Anninos93a}, these
data sets are interesting in their own right as dynamical black holes.
These black holes radiate both polarizations of the gravitational wave
field, and their horizon geometries can be so distorted they cannot be
embedded in a flat Euclidean space.  In future papers we will continue
to explore their dynamics in detail.  Moreover, just as in the
non-rotating case, these data sets can be used to understand the late
time behavior of coalescing, multiple black hole systems, as highly
distorted ``Kerr'' black holes will be formed in the process.  These
data sets can be thought of as representing initial conditions for the
late stages of that process, and should prove to be a valuable system
for studying it without having to evolve the orbits leading up to the
merger.

The paper is organized as follows: In section \ref{construction} we
detail the mathematical construction of the data sets and discuss some
of their properties, such as their masses, for several subclasses of
the spacetimes.  Then in section~\ref{sec:ah} we discuss a series of
tools we have developed to study the initial data sets, such as the
apparent horizons and their intrinsic geometry. In
section~\ref{survey} we survey many initial data sets, describe their
features, and discuss application of the hoop conjecture and Penrose
inequality.  Finally in section~\ref{summary} we summarize the
results.

\section{Constructing Distorted, Rotating Black Holes}
\label{construction}
\subsection{Initial Data}
\label{initial}
In this section we provide details of the construction of this new
family of distorted, rotating black hole data sets.  Some details were
provided in Ref.~\cite{Brandt94b}, but as this paper focuses
exclusively on the initial value problem we present a more
comprehensive treatment here.  To motivate the problem, consider first
the Kerr metric, written in Boyer-Lindquist form~\cite{Misner73}:
\begin{eqnarray}
g^{(K)}_{\mu\nu} &=& \left( \begin{array}{cccc}
g_{tt}{}^{(K)} & 0 & 0 & g_{t\phi}{}^{(K)} \\
0 & g_{rr}{}^{(K)} & 0 & 0 \\
0 & 0 & g_{\theta\theta}{}^{(K)} & 0 \\
g_{t\phi}{}^{(K)} & 0 & 0 & g_{\phi\phi}{}^{(K)}
\end{array} \right) \nonumber\\
g^{(K)}_{rr}& = &\rho^2/\Delta \nonumber\\
g^{(K)}_{\theta\theta}& = &\rho^2 \nonumber\\
g^{(K)}_{tt}& = &\left( a^2 \sin^2\theta-\Delta \right)/\rho^2 
\nonumber\\
g^{(K)}_{t\phi}& = &-2 a m r \sin^2\theta/\rho^2 \nonumber\\
g^{(K)}_{\phi\phi}& = &\left(\left(r^2+a^2\right)^2-
\Delta a^2 \sin^2\theta 
\right) \sin^2\theta/\rho^2 \nonumber\\
\Delta & = & r^2-2mr+a^2 \nonumber\\
\rho^2& = &r^2+a^2 \cos^2\theta ,\label{kerr metric}
\end{eqnarray}
where $m$ is the mass of the Kerr black hole, $a$ is the angular
momentum parameter, and a superscript ${}^{(K)}$ denotes the Kerr
spacetime.

We would like to put this in a form that is free of coordinate
singularities, and that has properties similar to the form used in the
non-rotating black hole
studies~\cite{Anninos93c,Bernstein94a,Bernstein93b}.  Generalizing the
transformation used in
Ref.~\cite{Anninos93c,Bernstein94a,Bernstein93b} we define a new
radial coordinate $\eta$ through
\begin{mathletters}
\begin{eqnarray}
r & = & r_+ \cosh^2\left(\eta/2\right)- r_- \sinh^2\left(\eta/2\right),\\
r_\pm & = & m \pm \sqrt{m^2-a^2}.
\end{eqnarray}
\label{Eqn:EtaCoordTransform}
\end{mathletters}

With this transformation the spatial part of the Kerr metric can be
written as
\begin{equation}
dl^2=\Psi_0^4 \left[e^{-2 q_0}\left(d\eta^2+d\theta^2 \right)
+\sin^2\theta d\phi^2 \right] \label{anametric}
\end{equation}
where
\begin{mathletters}
\begin{eqnarray}
\Psi_0^4&=&g^{(K)}_{\phi\phi}/\sin^2\theta \label{psiset}\\ 
\Psi_0^4 e^{-2 q_0}&=&g^{(K)}_{rr} \left( \frac{dr}{d\eta} \right)^2
=g^{(K)}_{\theta\theta}\label{q0set}.
\end{eqnarray}
\end{mathletters}\\
Thus we see that our coordinate
transformation~(\ref{Eqn:EtaCoordTransform}) has resulted in a
``quasi-isotropic gauge''~\cite{Evans86} for the Kerr spacetime.
Notice also that if $a=0$ then $q_0=0$ and we recover the
Schwarzschild 3-metric.  This metric is now in the form used in the
previous NCSA black hole
studies~\cite{Abrahams92a,Anninos93c,Bernstein94a,Bernstein94b}, although
here the functions $q_0$ and $\Psi_0$ are determined by the Kerr
spacetime.

One may check that the 3--metric defined by Eq.~(\ref{anametric}) is
invariant under the transformation $\eta \rightarrow -\eta$, i.e.,
there is an isometry operation across the throat of an Einstein-Rosen
bridge located at $\eta=0$, just as in the Schwarzschild
spacetime~\cite{Bernstein89}.  This construction has two geometrically
identical sheets connected smoothly at the throat, located at $\eta =
0$.  In terms of the more familiar radial coordinates, this condition
can be expressed as
\begin{equation}
\bar{r} \rightarrow \left(m^2-a^2\right)/
\left(4 \bar{r} \right),
\end{equation}
where $\bar{r}$ is a generalization of the Schwarzschild isotropic
radius, defined as
\begin{mathletters}
\begin{eqnarray}
\bar{r}&=&\frac{\sqrt{m^2-a^2}}{2}e^\eta\label{rbareta}.
\end{eqnarray}
$\bar{r}$ is related to the usual Boyer-Lindquist
radial coordinate via
\begin{eqnarray}
r&=&\bar{r} \left(
1+\frac{m+a}{2 \bar{r}}\right) \left(1+\frac{m-a}{2 \bar{r}}\right).
\label{rrbar}
\end{eqnarray} \label{rbar}
\end{mathletters} \\
Note that in the Kerr spacetime the horizon, located at
$r=m+\sqrt{m^2-a^2}$, is at $\bar{r}=\sqrt{m^2-a^2}/2$ in the
$\bar{r}$ coordinates, or at $\eta=0$, just as in previous studies of
the Schwarzschild spacetime~\cite{Bernstein89}.  The isometry
condition $\eta \rightarrow -\eta$ will be used in the construction of
more general initial data sets to be described below, and is also
imposed during the evolution as well~\cite{Brandt94c}.

The 3-metric, however, comprises only half of our initial data.  The
extrinsic curvature on the initial slice is required as well, and for the 
Kerr spacetime it is
given by the equations
\begin{mathletters}
\begin{eqnarray}
\hat K_{\eta\phi} &=& a m \left(2 r^2 \left(r^2+a^2\right)+
\rho^2 \left(r^2-a^2\right)\right) \sin^2\theta\,\rho^{-4} \\
 \hat K_{\theta\phi} &=& -2 a^3 m r \Delta^{1/2}
\cos\theta\,\sin^3\theta\,
\rho^{-4},
\end{eqnarray}
\label{kerrexcurv}
\end{mathletters}\\
where we note that $K_{ij}=\Psi_0^{-2} \hat K_{ij}$, and all extrinsic
curvature components excepting the two listed are zero.

One can show that this metric and extrinsic curvature satisfy the both
the Hamiltonian constraint
\begin{equation}
R+K^2-K^{ij} K_{ij}=0
\label{Hamiltonian}
\end{equation}
and the three momentum constraints
\begin{equation}
\nabla_i \left(K^{ij}-\gamma^{ij} K \right)=0,
\label{momentum}
\end{equation}
as they must.

Finally we must specify the lapse and shift to complete the system for 
evolution.  The choices that make the stationary nature of this Kerr spacetime
explicit are:
\begin{equation}
\frac{1}{\alpha^2} = 1 + 2\,m\,r \left( \frac{r^2+
	a^2}{\Delta \rho^2} \right) 
\end{equation}
and
\begin{equation}
\beta^{\phi} = \frac{-2 a m r}{\left(r^2+a^2\right)^2-a^2 \Delta \sin^2\theta}.
\end{equation}

We have now written the Kerr spacetime in 3+1 form, in a coordinate
system like that used to evolve the nonrotating distorted black holes
in previous work.  This is a stationary spacetime, and it can be
evolved using a dynamic slicing condition just as the Schwarzschild
spacetime has been evolved using maximal and other slicing
conditions. But just as in the case of Schwarzschild, we can distort
this rotating black hole with a gravitational (``Brill'') wave and study its
response.

We accomplish this by adding a free function $q(\eta,\theta)$ to the
conformal 3--metric in Eq.~(\ref{anametric}), just as in
Ref.~\cite{Bernstein94a}.  Then the conformal factor $\Psi_0$, which
is a solution only for the stationary Kerr spacetime, will no longer
satisfy the Hamiltonian constraint~(\ref{Hamiltonian}).  Instead we
must solve this constraint equation for the general conformal factor
$\Psi$ as we discuss below.  This generalization results in a 3-metric
with the following form:
\begin{equation}
dl^2 = \Psi^4 \left[ e^{2\left(q-q_0\right)}\left(d\eta^2+d\theta^2\right)
+\sin^2\theta d\phi^2 \right].\label{3metric}
\end{equation}
Apart from the function $q_0$ this is the same 3-metric used for the
non-rotating distorted black hole spacetime.  This illustrates the
fact that unlike the Schwarzschild spacetime the undistorted Kerr
spacetime is {\em not} conformally flat ($\hat R \neq 0$, where $\hat
R$ is the scalar curvature of the conformal part of the 3--metric).
We may, in fact, make the spacetime conformally flat by adding a
``Brill wave''~\cite{Bowen80} specified by $q=q_0$.  This distorts the
Kerr black hole, which breaks the stationary nature of the spacetime.
As we will show below, appropriate choices of this function $q$ (along
with appropriate solutions to the momentum constraint for the
extrinsic curvature terms) can lead also to Schwarzschild, the NCSA
distorted non-rotating black hole (as in Ref.~\cite{Bernstein94a}),
the Bowen and York rotating black hole~\cite{Bowen80}, or a distorted
Bowen and York black hole.

The function $q$, representing the
Brill wave, can be chosen somewhat arbitrarily, subject to symmetry
conditions on the throat, axis and equator, and falloff conditions at
large radii~\cite{Bernstein94a,Brandt94b}.  The function $q$
will be chosen to have an inversion symmetric (i.e., symmetric under
$\eta \rightarrow -\eta$) Gaussian part given by:
\begin{mathletters}
\begin{eqnarray}
q &=& \sin^n\theta\, q_{G}, \\
q_{G} & = & Q_0 \left( e^{-s_+}+e^{-s_-} \right), \\
s_\pm & = & \left(\eta \pm \eta_0 \right)^2/\sigma^2.\label{metric form}
\end{eqnarray}\label{brill}
\end{mathletters} \\
This form of the Brill wave will be characterized by several
parameters: $Q_0$ (its amplitude), $\sigma$ (its width), $\eta_0$ (its
coordinate location), and $n$, specifying its angular dependence,
which must be positive and even.  We note that Eqs.~(\ref{brill})
simply provide a convenient way to parameterize the initial data sets,
and to allow us to easily adjust the ``Brill wave'' part of the
initial data.  Many other devices are possible.  To narrow the field
of study, in this paper we restrict ourselves to sets characterized by
 $n=2$, and $\eta_0=0$ for all data sets, and will commonly use
$\sigma=1.5$

With this form of the Brill wave function $q$, a number of black hole
data sets can be constructed.  As discussed above, given the
appropriate extrinsic curvature choice, the stationary Kerr solution
results if $q=0$.  For other choices the Bowen and York form results
if $q=q_0$, a {\em distorted} Bowen and York spacetime can be made by
setting $q=\sin^n\theta q_{G}+q_0,$ and a distorted Kerr spacetime can
be made by setting $q=\sin^n\theta q_{G}$.  Furthermore, for
spacetimes without angular momentum this metric form contains both the
NCSA Brill wave plus black hole spacetimes and, as we will see below,
a new black hole spacetime corresponding to odd-parity distortions of
the Schwarzschild spacetime.

As a consequence of this generalization of the metric, we must now solve
both the Hamiltonian and momentum constraints.  The Hamiltonian 
constraint equation~(\ref{Hamiltonian}) can be expanded in coordinate
form to yield
\begin{eqnarray}
\frac{\partial^2 \Psi}{\partial\eta^2}&+&\frac{\partial^2 
\Psi}
{\partial\theta^2}+ 
\frac{\partial \Psi}{\partial\theta}\cot\theta -\frac{\Psi}{4} = \nonumber\\
&-&\frac{\Psi}{4} \left( 
\frac{\partial^2}{\partial\eta^2}\left(q-q_0\right)+
\frac{\partial^2}{\partial\theta^2}\left(q-q_0\right)
\right)\nonumber\\
&-&\frac{\Psi^{-7}}{4}\left( \hat{H}_E^2 \sin^2\theta
+\hat{H}_F^2 \right),
 \label{Energy Constraint}
\end{eqnarray}
where we have assumed the initial slice is maximal, i.e. tr$K=0$.
There are in principle three momentum constraints to be satisfied 
by the extrinsic curvature components.  However, in our spacetimes
we require a ``time-rotation'' symmetry about the initial slice, or 
invariance of the metric under $(t,\phi) \rightarrow (-t,-\phi)$.
Writing the extrinsic curvature tensor as
\begin{eqnarray}
K_{ij} & = & \Psi^{-2} \hat{H}_{ij}\nonumber\\
&=& \Psi^{-2} \left( \begin{array}{ccc}
\hat{H}_A & \hat{H}_C & \hat{H}_E \sin^2\theta \\
\hat{H}_C & \hat{H}_B & \hat{H}_F \sin\theta \\
\hat{H}_E \sin^2\theta & \hat{H}_F \sin\theta & \hat{H}_D \sin^2\theta
\end{array} \right),  
\end{eqnarray}
this time-rotation symmetry requires that
$\hat{H}_A=\hat{H}_B=\hat{H}_C=\hat{H}_D=0$, leaving only $\hat{H}_E$
and $\hat{H}_F$ to be determined.  This also automatically
makes the initial slice maximal.

Under these conditions the only nontrivial component
of the momentum constraints is the $\phi$ component:
\begin{equation}
\partial_\eta \hat{H}_{E} \sin^3\theta+
\partial_\theta \left( \hat{H}_F \sin^2\theta \right) =0.
\label{mom constraint}
\end{equation}
Note that this equation is independent of the functions $q$ and $q_0.$
This enables us to choose the solutions to these equations independently
of our choice of metric perturbation.

At this point we may regard Eqs.~(\ref{3metric}-\ref{mom constraint})
as defining the problem we wish to solve.  The boundary conditions for
these equations, which are extensions of those used in
Ref.~\cite{Bernstein94a}, may be summarized as:\\

{\em Isometry Conditions:} Across the throat of the black hole,
labeled by $\eta=0$, we can demand a condition that the spacetime has
the same geometry as $\eta\rightarrow-\eta$. We can require this not
only for the initial data, but also throughout the evolution of the
spacetime. This condition may be stated in one of two ways (in
axisymmetry) to allow for different slicing conditions.  Each choice
must result in a symmetric $K_{\eta\phi}$ and an anti-symmetric
$K_{\eta\theta}$ to be consistent with the Kerr solution.  Thus, the
form of the isometry condition must be different for the symmetric and
anti-symmetric lapse spacetimes: $\eta \rightarrow -\eta$ for the
former, $(\eta,\phi)\rightarrow (-\eta,-\phi)$ for the latter.

If a lapse that is anti-symmetric across the throat is desired, the 
metric elements with a single $\eta$ index are anti-symmetric across 
the throat, while those with zero or two indices are symmetric. The 
extrinsic curvature components have the opposite symmetry of their 
corresponding metric elements. The shift $\beta^\eta$ is anti-symmetric 
across the throat, while all other shifts are symmetric. 

If a symmetric lapse is desired, the metric elements with a single
$\eta$ index or single $\phi$ index (but not both) will be
anti-symmetric at the throat and all others will be symmetric.
Extrinsic curvature components will have the same symmetries as their
corresponding metric elements. The $\beta^\eta$ and $\beta^\phi$
shifts will be anti-symmetric, and the $\beta^\theta$ shift will be
symmetric.  With these symmetries enforced, the initial data and all
subsequent time slices will be isometric across the throat.  One can
verify that all Einstein equations respect these symmetries during the
evolution if they are satisfied initially.\\ {\em Symmetry Axis:} All
metric elements, extrinsic curvature components, and shift components
with a single $\theta$ index are anti-symmetric across the axis.  The
remainder are symmetric.\\ {\em Equatorial Plane:} At the equator
there are two possible symmetries, the Kerr symmetry and the ``cosmic
screw'' symmetry (see section~\ref{cscrew} below).  For the Kerr
symmetry, $\theta \rightarrow \pi-\theta$, all metric components,
extrinsic curvature components or shifts with a single $\theta$ index
are anti-symmetric.  The remainder are symmetric. For the cosmic-screw
type boundary conditions the symmetry at the equator is
$(\theta,\phi)\rightarrow(\pi-\theta,-\phi)$ and those metric
elements, extrinsic curvature components, or shift components that
contain one $\theta$ or one $\phi$ index (but not both) are
anti-symmetric.  The remainder are symmetric.\\ {\em Outer Boundary:}
At the outer boundary a Robin condition is used for $\Psi$. This
condition gives the correct asymptotic behavior in the conformal
factor to order $r^{-2}$.  For the metric given in the form
(\ref{3metric})  $\Psi$ has the form $\Psi = e^{\eta/2}+ (m/2)
e^{-\eta/2}+...$ and therefore obeys the differential equation $\Psi+2
\partial_\eta \Psi = 2 e^{\eta/2}$.  The conformal factor is always
symmetric at the throat, axis, and equator.

With this description of the problem and boundary conditions, we
summarize a number of families of rotating black hole solutions that
are possible in this formulation:
\subsubsection{Kerr and Distorted Kerr}
As shown above, Eqs.~(\ref{3metric}-\ref{mom constraint}), together
with the boundary conditions, are satisfied by the stationary Kerr
spacetime with the
transformations~(\ref{Eqn:EtaCoordTransform}-\ref{kerrexcurv}).  In
Eq.~(\ref{3metric}), this spacetime corresponds to $q=0$, but as we
have mentioned above, we may add a Brill wave function $q=\sin^n\theta
q_{G}$ to distort the Kerr spacetime.  From the standard ADM
expressions for the mass and angular momentum of the
spacetime~\cite{Omurchada74} we have
\begin{mathletters}
\begin{eqnarray}
M_{ADM} &=& -\frac{1}{2\pi}\oint_S \nabla_a (\Psi e^{-\eta/2}) dS^a
\label{admmass} \\
P_a &=& \frac{1}{8\pi} \oint_S
\left(H_a^b-\gamma_a^b H\right) dS_b,
\label{admp}
\end{eqnarray}
\end{mathletters} \\
or in terms of the variables defined in this paper,
\begin{mathletters}
\begin{eqnarray}
 M_{ADM} & = & \int_0^\pi e^{\eta/2} \left(
-\partial_\eta \Psi+
\Psi/2 \right) \sin\theta d\theta \label{admmass2},\\
J & = & P_{\phi} = \frac{1}{4} \int_0^\pi \Psi^6 H_E
\sin^3\theta d\theta.
\label{angularmom}
\end{eqnarray}
\end{mathletters} \\
If one evaluates these expressions for the undistorted Kerr spacetime,
($q=0$), one recovers the expected results $M_{ADM} = m$ (which should
be evaluated at $\eta=\infty$) and $J = a\,m$ (which may be evaluated
at any radius in axisymmetry).  In practice, we evaluate $m$ at the
outer edge of our grid, and $J$ is simply an input parameter.  We have
evaluated and verified $J$ at various radii, as reported in previous
work
\cite{Brandt94b}.

\subsubsection{Distorted Bowen and York}
Perhaps the simplest nontrivial choice of solution for the momentum
constraint is given by the Bowen and York~\cite{Bowen80} solution
\begin{mathletters}
\begin{eqnarray}
\hat H_E&=&3 J \\
\hat H_F&=&0, \label{B and Y solution}
\end{eqnarray}
\end{mathletters}
where $J$ is the total angular momentum of the spacetime.  This solution
clearly satisfies the momentum constraint.  In the language of this
paper, the original Bowen and York
solution has a Brill wave function determined by
\begin{equation}
q=q_0,
\end{equation}
i.e., it is conformally flat, but we may distort it by setting
\begin{equation}
q=q_0 + \sin^n\theta q_{G}.
\end{equation}
This is the solution we often use in constructing distorted rotating
black hole spacetimes, and a number of them have been evolved and
discussed in Ref.~\cite{Brandt94c}.  It is somewhat simpler to
construct than the distorted Kerr because we can choose our total
angular momentum and distortion parameters without needing to know the
ratio $J/m^2$ (an additional value that we would need to initialize
the Kerr momentum solution).  Knowing this parameter would require us
to know the value $m$, and we will not know that parameter until after
we solve the Hamiltonian constraint.  It is possible, in
principle, to iterate between guesses for the parameter $J/m^2$ and
solutions for the Hamiltonian constraint until arriving at a correctly
initialized Kerr momentum solution.

We noted above that the Kerr solution is not conformally flat, while
the Bowen and York solution is conformally flat.  Therefore one can
consider adding a Brill wave to the Kerr solution, making it a
conformally flat Bowen and York solution.  One can study the form of
the wave functions $q$ and $q_0$ to understand what sort of Brill wave
is required to ``flatten'' the Kerr solution.  By equating the series
expansions of $q_0$ and our Brill wave function
(c.f. Eq. (\ref{q0set}))
\begin{equation}
q_{{\rm approx}} = a_1 \sin^2\theta e^{-\eta^2/\sigma_1^2}+
	a_2 \sin^4\theta e^{-\eta^2/\sigma_2^2}.
\end{equation}
 to second order in $\eta$ and $\theta$, one obtains relationships
 between the wave shape parameters $a_1$, $a_2$, $\sigma_1$, and
 $\sigma_2$ and the Kerr parameters $a$ and $m$.  It turns out that for
 $a/m < .9$ we find that the $\sin^2\theta$ term dominates the
 $\sin^4\theta$ term, and $\sigma_1$ varies in value from $1.26$ to
 $1.7$ for rotation parameters $0.01 \leq a/m \leq 0.8$.  Furthermore
the amplitude $a_1$ varies smoothly from zero to about $0.38$ for all
rotation parameters $0 \leq a/m \leq 1$.  We conclude that the Bowen
and York distortion is similar to a low amplitude $n=2$, $\sigma =1.5$
Brill wave distortion on a Kerr black hole, which is quite similar in
shape and size to the distorted, non-rotating black holes evolved
previously~\cite{Abrahams92a}.

To get a feel for the difference between the $q_0$ function and its
approximation, $q_{approx}$, we calculated the ADM value of
 $a/M_{ADM}$ as we would for the Kerr spacetime, except
that we used $q_{approx}$ instead of $q_0$ in the metric.  We found
that the error in $a/M_{ADM}$ was less than $1\%$, even for $a/m=0.9$.

\subsubsection{Odd-parity Distorted Schwarzschild}
Odd-parity gravitational modes, as defined by Regge and
Wheeler~\cite{Regge57}, are metric perturbations with parity
$(-1)^{\ell+1}$ where $\ell$ is the angular index of the tensor
spherical harmonic describing the wave.  These modes, in
axisymmetry, are formed only from perturbations in $g_{r\phi}$ and
$g_{\theta\phi}$.  The evolution of these modes does not couple
linearly to the even-parity modes, and if there is no odd-parity
disturbance on the initial slice of the spacetime these modes are never
excited even when the full non-linear equations are used.

We can create odd-parity distorted non-rotating spacetimes simply
by finding solutions to the momentum constraint equation~(\ref{mom
constraint}) which drive the evolution of $g_{\theta\phi}$, which
satisfy the boundary conditions, and yield $J=0$ when the integral
given in Eq.~(\ref{angularmom}) is performed.

As a framework for momentum solutions, we consider the simple class
defined by:
\begin{mathletters}
\begin{eqnarray}
\hat H_E &=& f_1(\theta)+f_2(\eta) \left(
4\,\cos\theta\,f_3(\theta)+\sin\theta\,\partial_\theta
f_3(\theta)\right) \\
\hat H_F &=& -\partial_\eta f_2(\eta) \sin^2\theta\, f_3(\theta).
\end{eqnarray}\label{theform}
\end{mathletters}\\
This solves the momentum constraint where $f_1$, $f_2$ and $f_3$ are
arbitrary functions which satisfy the following symmetries:
\begin{mathletters}
\begin{eqnarray}
f_1(\theta) &=& f_1(-\theta) = f_1(\pi-\theta)\\
f_3(\theta) &=& -f_3(-\theta) = -f_3(\pi-\theta)\\
f_2(\eta) &=& f_2(-\eta) \\
\lim_{\eta \rightarrow \infty} f_2(\eta) &=& 0 
\end{eqnarray}
\end{mathletters}\\
The Kerr solution is not a member of this class, but the Bowen and
York solution does fit within it ($f_1 = 3 J$, $f_2 = f_3 = 0$).

Note that any spacetime with $f_1=0$ will have zero angular momentum,
as $\hat H_E$ falls off for large values of the radial coordinate
$\eta$, and furthermore the angular part of the integral in
Eq.(\ref{angularmom})is identically zero by construction.  We require
$f_2$ to go to zero only so that we can make the spacetime
asymptotically Kerr.

We now turn to a solution that has no net angular momentum, but one
that is not time symmetric about the initial slice.  This solution
corresponds to a ``Schwarzschild'' black hole with both even- and
odd-parity distortions.  This generalizes the NCSA Brill wave plus
black hole spacetimes~\cite{Bernstein94b}, which allowed only
even-parity distortions.

For this we choose the extrinsic curvature components such that
$f_1 = 0$, $f_2=q_G^\prime$, and $f_3
=\cos\theta\,\sin^{n^\prime-3}\theta$.  In order to obey symmetry and
regularity conditions, $n^\prime$ must be odd and have a value of at
least 3 (This was arranged to be suggestive of a radiation $\ell$-mode.
Using $n=3$ gives almost pure $\ell=3$, while using $n=5$ gives almost
a pure $\ell=5$ component~\cite{Brandt94c}.).

As this data set contains the odd-parity polarization of the
gravitational wave field, but no angular momentum, it is simpler to
evolve than than the rotating black holes described above, but has
richer physical content than the standard Brill wave plus black hole
data sets.  A few of these data sets have been evolved and discussed in
Ref.~\cite{Brandt94c}, where both even- and odd-parity waveforms were
extracted.

\subsubsection{Cosmic Screws}
\label{cscrew}
There is another solution related to the odd-parity distorted
Schwarzschild data set discussed above.  While that data set has
equatorial plane symmetry, we can alternatively construct a solution
that has something like an opposite sense of rotation below the
equatorial plane.  As discussed above, this symmetry is $(\theta,\phi)
\rightarrow (\pi-\theta,-\phi)$, thus requiring $\hat H_E$ to be
antisymmetric and $\hat H_F$ to be symmetric across the equator.

Axisymmetric data sets describing the collision of two
counter-rotating black holes have been dubbed ``cosmic screws.''
As our data sets contain only one black hole, it can be considered to
represent a later stage of evolution of cosmic screws, where the holes
have already merged.  Such data sets should be useful in studying the
late time behavior of colliding spinning black holes, and will be
evolved in a future paper.  The formalism of the previous section
carries over, but now the symmetry conditions on the equator are
reversed: $f_1(\theta) = - f_1(\pi-\theta)$ and $f_3(\theta) =
f_3(\pi-\theta)$.

With these considerations in mind, we see that $f_1 = 3 J \cos\theta$,
and $f_2 = f_3 = 0$ is suggestive of a cosmic screw spacetime (in this
case we do not claim that $J$ is an angular momentum, but use this
notation because it is suggestive of the angular momentum of one of
the colliding black holes in a cosmic-screw spacetime).

We can also choose $f_1 = 0$, $f_2 = q_G^\prime$, and $f_3 =
\sin^{n^\prime-2}\theta$ which is similar to our odd-parity distortion
of Schwarzschild except that the equatorial boundary conditions are
like those of the cosmic screw.  Here $n^\prime$ should be even
and have a value of at least 2 in order to obey symmetry and
regularity requirements.

\subsection{Numerical Techniques}
 To solve the Hamiltonian constraint equation we use a Newton-Raphson
iteration procedure.  The conformal factor in the constraint equation
is linearized about a trial function.  The initial trial function,
$\Psi_0$, is usually chosen to be $2\,{\rm cosh}(\eta/2)$, because
that is the solution for a spherical black hole with mass $M=2$.  At a
given iteration $i$ we assume the solution has the form
$\Psi=\Psi_i+\delta \Psi$, where $\delta\Psi$ is a small correction
function.  We then evaluate the residual $\epsilon_i$ of this guess
for $\Psi_i$,
\begin{mathletters}
\begin{eqnarray}
\epsilon_i &=& \bigtriangleup \Psi_i+\frac{\Psi_i}{4} \left(\partial^2_\eta
\left(q-q_0 \right)+\partial^2_\theta \left( q-q_0 \right) \right)\nonumber\\
&&-\frac{\Psi_i^{-7}}{4} \left( \hat H_E^2 \sin^2\theta+\hat H_F^2 
\right)
\label{step 1}
\end{eqnarray}
and use it as a source term for the correction function $\delta \Psi$
\begin{eqnarray}
\bigtriangleup \delta \Psi&+&\frac{\delta \Psi}{4}
\left(\partial^2_\eta
\left(q-q_0 \right)+\partial^2_\theta \left( q-q_0 \right) \right)\nonumber\\
&+&7 \delta \Psi \frac{\Psi_i^{-8}}{4} \left( \hat{H}_E^2
\sin^2\theta+\hat{H}_F^2 \right) = -\epsilon_i. \label{step 2}
\end{eqnarray}
\end{mathletters}
Eq.~(\ref{step 1}) is used merely to evaluate $\epsilon_i$, but
Eq.~(\ref{step 2}) is solved with a numerical elliptic solver for
 $\delta\Psi$.  The boundary conditions on the grid are given as
follows: $\delta \Psi$ is symmetric across the throat, the axis, and
the equator.  At the outer boundary we use the Robin condition,
 $\partial_\eta \delta \Psi = -\frac{1}{2} \delta \Psi$.  Typically we
use $\epsilon = 10^{-7}$ as our stopping criterion.

Once $\delta \Psi$ is obtained, we try again, with
$\Psi_{i+1}=\Psi_i+\delta\Psi$.  We repeat this
procedure until the a solution to Eq. (\ref{Energy Constraint}) with
appropriately small error (maximum value of $\epsilon_i$ on the grid)
is achieved~\cite{Cook90}.

Some of the data sets considered in this paper use Brill waves that
are very wide.  Because of this, one may worry whether the outer
boundary of the grid is sufficiently large to get an appropriate
asymptotic behavior for the Hamiltonian constraint.  To determine whether
the position of $\eta_{max}$ is valid, we measure the quantity
\begin{equation}
\xi = \partial_\eta{\rm log}\left|\Psi-e^{\eta/2}\right|
\end{equation}
which should approach $-1/2$ in a spacetime that asymptotically
approaches the Kerr spacetime.  We adopt the criterion that $\xi$ must
be equal to this value to within $0.1\%$.  If it is not, we regard our
choice for value of $\eta_{max}$, the outer boundary of our grid, to
be too small and adjust accordingly.

\subsection{Limits on the magnitude of Brill wave amplitudes}
\label{bwlimits}

Cantor and Brill~\cite{Cantor81} showed that for a certain class of
spacetimes there are limits on the value of the amplitude of the Brill
wave in a spacetime with topology $R^3$.  Their basic technique can be
applied to these rotating black hole spacetimes, which have topology
$R\times S^2$.  We present a simplified version of this analysis.

The strategy is to derive an inequality that contains an arbitrary
function $f$, but does not explicitly contain the conformal factor
$\Psi$, and which requires asymptotic flatness of the spacetime as one
of its conditions for validity.  We use the freedom in $f$ to
locate Brill wave parameter spaces which fail to satisfy the
inequality.  When the inequality fails, it is then clear that the
assumption of asymptotic flatness has been violated, and the value of
the input Brill wave parameters which specify it are outside our
domain of interest.

For this section, we define the metric as
\begin{equation}
ds^2 = \tilde \Psi^4 e^{2 \eta} \left[
e^{2\left(q-q_0 \right)} \left( d\eta^2+d\theta^2 \right)+
\sin^2\theta\,d\phi^2 \right].
\end{equation}

Brill started by considering the volume integral
\begin{equation}
\int \hat\nabla f \cdot \hat\nabla f\, dV
\end{equation}
where $f = u \tilde \Psi$, and where $u$ or $f$ is an arbitrarily
chosen function.  We can expand the above integral for $f$ in terms of
$u$ and $\tilde \Psi$: 
\begin{equation}
\hat \nabla \cdot \left( u^2 \tilde \Psi \hat \nabla \tilde \Psi \right) -
u^2 \tilde \Psi \hat \triangle \tilde \Psi =
2 u \tilde \Psi \hat \nabla u \cdot \hat \nabla \tilde \Psi+
u^2 \hat \nabla \tilde \Psi \cdot
\hat \nabla \tilde \Psi.
\end{equation}
We have now obtained both a surface and a volume piece for the
integral.  Because we want a final equation which does not include
$\tilde \Psi$, we discard the terms under the integral which still
contain it explicitly after our transformation.  Since these terms are
manifestly positive, the remainder must be less than our original
integral.  We now have an inequality.  Next we require that
$\tilde\Psi$ not vanish (this is only necessary if we have
non-zero extrinsic curvature), is finite, and that $\hat \nabla u$ go
to zero faster than $r^{-3/2}$.

To evaluate the surface terms we make use of the asymptotic properties
of $\tilde \Psi$, namely
that $\tilde \Psi = 1 + \frac{m}{2 r}+O(r^{-2})$ at large radii.  The
surface term is
\begin{equation}
\oint
 u^2 \tilde \Psi \hat \nabla \tilde \Psi \cdot dS.
\end{equation}
This last expression will be $-2 \pi$ times the ADM mass of the system
if we choose $u \rightarrow \Psi^{-1}$ at $\eta \rightarrow \infty$.  If we choose $u
\rightarrow 0$ as $\eta \rightarrow \infty$ then the integral will not
evaluate to the ADM mass (in fact it will obtain zero).  When we
perform this surface integral at the throat of our spacetime we
exploit the isometry
\begin{equation}
\partial_\eta \left(\tilde \Psi e^{\eta/2} \right) = 0,
\end{equation}
to obtain the final form for our inequality
\begin{eqnarray}
2 \pi f_{\eta=\infty}^2 M_{ADM} &+&\int \left( \hat\nabla f \cdot \hat\nabla f
+\frac{1}{8} f^2 \hat R \right)dV >\nonumber\\
&&\pi \int_0^\pi
f_{\eta=0}^2 \sin\theta\, d\theta.\label{ineq}
\end{eqnarray}
where we have assumed that $f$ is independent of $\theta$ at large $\eta$.

This inequality can only fail if our assumptions about the form of
$\tilde \Psi$ are incorrect.  Therefore, when it does fail, we can
conclude that an asymptotically flat metric cannot be formed by
the given Brill wave parameters.

We explored the regime where $q = \sin^n\theta\,q_G+q_0 = 2 Q_0
\sin^2\theta\,e^{-\eta^2/\sigma^2}$ and found two forms for
$f$ which provided us with good bounds on the integral.  These forms
are
\begin{equation}
f={\rm exp}\left(-\epsilon \eta^2-\mu \eta/2\right)\label{func0}
\end{equation}
and
\begin{equation}
f=\cos^4\theta\,{\rm exp}\left(-\epsilon \eta^2-
\mu \eta/2\right).\label{func4}
\end{equation}
Replacing the inequality with an equality in Eq.~(\ref{ineq}),
performing the integral, and then solving for $Q_0$ provides us with a
function, $Q_0(\mu,\epsilon,\sigma)$, whose range supplies the
forbidden amplitude values $Q_0$ and whose domain is $\epsilon > 0$,
 $\sigma> 0$, and $\mu$ is a real number.

\begin{figure}
\epsfxsize=225pt \epsfbox{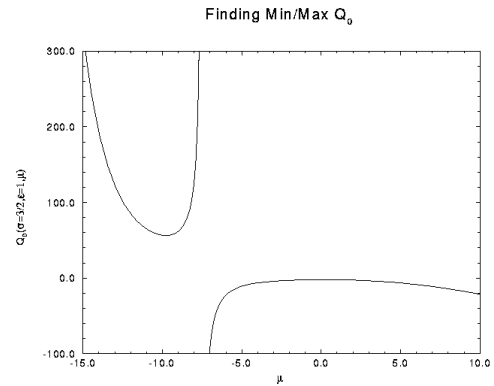}
\caption{
We plot the range of the Brill wave amplitude function $Q_0$which are
inconsistent with asymptotically flat solutions of the Hamiltonian
constraint for a given ``trial function'' (see text).  In this example
plot, the values outside the range $-2.317 < Q_0 < 56.24$ are
excluded.}
\label{fig:twoextrema}
\end{figure}

The functions $Q_0(\mu,\epsilon,\sigma)$ that we obtained were able to
provide both an upper and a lower bound for the allowed values of the
input parameter $Q_0$ because they trace out a function similar to an
hyperbola in $\mu$ space for given values of $\sigma$ and $\epsilon$.
The behavior of one function is illustrated in
Fig.~\ref{fig:twoextrema}.  In this figure we see two distinct
extrema.  One extremum is a greatest possible negative value for $Q_0$
and provides us with a lower bound for this parameter.  The other is a
minimum in the possible positive values for $Q_0$ and this provides us
with a maximum value for $Q_0$.

In Table~\ref{tab:upper} we have give the strictest upper bound
for the function $Q_0$ that we were able to find using the above
inequality, Eq.~(\ref{ineq}), and trial functions,
Eqs.~(\ref{func0},\ref{func4}). These bounds, together with the
parameters $\mu$ and $\epsilon$ which were chosen to minimize them,
are given in the table.

\vbox{
\begin{table}
\begin{tabular}{|d||d|d|c|d|}
$\sigma$ & $\epsilon$ & $\mu$ & $n$ & $Q_0$ \\ \hline \hline
0.25 & 1.726  & -2.928 & 0 & 3.283 \\ \hline
0.50 & 1.078  & -3.186 & 0 & 5.122 \\ \hline
0.75 & 0.8455 & -3.64  & 0 & 7.516 \\ \hline
1.00 & 0.7042 & -4.037 & 0 & 10.74 \\ \hline
1.25 & 0.6171 &  1.096 & 4 & 11.97 \\ \hline
1.50 & 0.5580 &  1.097 & 4 & 10.99 \\ \hline
1.75 & 0.5049 &  1.087 & 4 & 10.35 \\ \hline
2.00 & 0.4587 &  1.076 & 4 & 9.914 \\ \hline
3.00 & 0.3300 &  1.046 & 4 & 8.993 \\ \hline
100.0 & 0.0100 & 1.014 & 4 & 7.6113 \\ \hline
\end{tabular}
\caption{Upper limits on $Q_0$.  The parameters $\epsilon$, $\mu$, and
 $n$ are arbitrary parameters were chosen in an attempt to find the
smallest upper bound for the allowed values of $Q_0$, the wave
amplitude, as a function of the wave width, $\sigma$.}
\label{tab:upper}
\end{table}
}

If one explores the $\mu=1$ form of this distortion and takes the
limits $\sigma \rightarrow \infty$ and $\epsilon \rightarrow 0$ in
that order one arrives at the amplitude $Q_0 = 7.577$.  Thus, it seems
that while there are more strict limits on larger width waves in
general, there is no reason to suspect that any width is too large.

In Table~\ref{tab:lower} we have the strictest lower bound
for the function $Q_0$ that we were able to find using the methods of
this section.

\vbox{
\begin{table}
\begin{tabular}{|d||d|d|d|d|}
$\sigma$ & $\epsilon$ & $\mu$ & $n$ & $Q_0$ \\ \hline \hline
0.25 & 0.0025 & 2.81 & 0 & -1.261 \\ \hline
0.50 & 0.0025 & 2.31 & 0 & -1.437 \\ \hline
0.75 & 0.0558 & 1.94 & 0 & -1.486 \\ \hline
1.00 & 0.1053 & 1.62 & 0 & -1.456 \\ \hline
1.25 & 0.1133 & 1.44 & 0 & -1.404 \\ \hline
1.50 & 0.1081 & 1.33 & 0 & -1.351 \\ \hline
1.75 & 0.1001 & 1.26 & 0 & -1.303 \\ \hline
2.00 & 0.0921 & 1.21 & 0 & -1.260 \\ \hline
100.0 & 0.0487 & 1.00 & 0 & -0.765 \\ \hline
\end{tabular}
\caption{Lower limits on $Q_0$.  The parameters $\epsilon$, $\mu$, and
 $n$ are arbitrary parameters were chosen in an attempt to find the
greatest lower bound for the allowed values of the wave amplitude
 $Q_0$, and the wave width, $\sigma$.}
\label{tab:lower}
\end{table}
}

It is interesting to note, however, that the pure Bowen and York
spacetime (which is parameterized by $J$, not $Q_0$ and can be
regarded as a distortion of the Kerr spacetime) can be specified
apparently without limit.  This is consistent with the calculation
done here, because the conformal Ricci scalar vanishes for conformally
flat spacetimes such as the Bowen and York spacetime.  This assures us
that the inequality given in Eq.~(\ref{ineq}) is trivially satisfied.
Numerical evaluations of the Bowen and York black hole (see
~\cite{Choptuik86b}) seem to get asymptotically closer to the extremal
Kerr spacetime as $J$ is increased, i.e. $J/m^2 = a/m$ gets ever
closer to unity confirming this observation.

We note also that we can use Eq.(\ref{ineq}) to obtain a lower bound
on the ADM mass for numerically derived data sets.  Using $f=1$ we obtain
\begin{equation}
M_{ADM} >  1+\frac{1}{4} \int_0^\pi \left(q-
q_0 \right) \sin\theta\,d\theta\label{Eqn:Bound}
\end{equation}
where the integral is evaluated on the throat.  We calculate this
after computing $\Psi$ on the initial slice to determine whether the
ADM mass of the spacetime is consistent with asymptotic flatness.

\section{Apparent Horizons}
\label{sec:ah}
\subsection{The numerical method}
We will use the same method for finding the AH (apparent horizon) as
described in previous work~\cite{Bernstein93a,Brandt94c}, where
surfaces that satisfy the trapped surface equation
\begin{equation}
D_i s^i+K_{ij} s^i s^j-K = 0
\end{equation}
are sought.  In the above, $s^i$ is an outward pointing unit normal
vector which defines the surface. For initial data sets of the type
considered in this paper, the terms containing the extrinsic curvature
drop out, and therefore finding the apparent horizon reduces to
finding an extremal surface.  We will, however, be using a different
strategy for searching than we used when we were evolving our
spacetimes~\cite{Brandt94b,Brandt94c}.  For extremely distorted data
sets we found that multiple trapped surfaces may exist, and it is
important to find the {\em outermost} surface.  Therefore, we use a
``brute force'' approach and repeatedly call the solver, using many
constant radial zones covering the inner portion of the grid as an
initial guess.  After the horizon finder has finished executing for
each initial guess, the solution with the largest radial coordinate
position that satisfied the tolerance of the solver was determined to
be the apparent horizon.

This strategy relies on our previous experience, which has shown us
that even horizons with highly distorted geometry are nearly spherical
in coordinate space, in our choice of coordinates.  Moreover, we know,
from Gibbons~\cite{Gibbons72}, that our apparent horizon must have the
topology of a 2-sphere.  As in previous
papers~\cite{Brandt94c,Anninos93a}, we will study the intrinsic
geometry of the black hole horizon as a measure of its distortion.

\subsection{Measures of the horizon}
\label{tools}

There are a number of geometric quantities that can be used to measure
the apparent horizon that provide insight to the physical
characteristics of the black hole.  Examples are its shape, area, and
local curvature. One straightforward measure of a black hole's shape
in axisymmetry is the ratio of its polar circumference, $C_p$, to its
equatorial circumference, $C_e$.  This ratio has been used previously,
but we point out that it has an inadequacy.  If used to study the
surface of an event horizon, this number may be infinitely large for
some configurations; one need merely consider the case where two equal
mass black holes are colliding and are just touching so that $C_e$
vanishes. Although we do not expect this problem to affect the
apparent horizon, however, we would like to use the same measure of
oblateness for both apparent and event horizons, one which better
captures the intuitive concept of oblateness.  One such measure is
$C_p/C_A$, where $C_A = 2 \pi (2 \sqrt{A/16 \pi})$ is the ``areal
circumference.''  Note that the quantity $C_p/C_A$ tends to be much
smaller than $C_p/C_e$.  For the case of a $\sigma=3.5$, $Q_0=5.64$,
black hole apparent horiozon we have $C_p/C_A=85$ and
$C_p/C_e=19\times 10^3$.

Another important quantity that can be obtained from the horizon is
its mass
\begin{equation}
M_{AH}^2 = \frac{A}{16\pi}+\frac{4 \pi J^2}{A},
\end{equation}
and the related
quantity, the MRL (maximum radiation loss),
\begin{equation}
MRL=1-M_{AH}/M_{ADM}.
\end{equation}
The MRL is the amount of energy that can be emitted if the black hole
swallows none of the radiation energy present outside the horizon on
the initial slice during its subsequent evolution.

We will also be interested in discussing what we call the ``transition
points'' of the horizon.  These are points in parameter space of the
initial data problem at which the apparent horizon ceases to be on the
throat (which is always a trapped surface) of the spacetime. The
transition points can be accompanied by discontinuous changes in the
character of the horizon, as measured by the quantities discussed here.

\section{Survey of Initial Data Sets}
\label{survey}
We now apply a number of these horizon measures developed above to
analyze a series of initial data sets, beginning with distorted,
non-rotating holes.

\subsection{Time-symmetric Brill wave data}
\begin{figure}
\epsfxsize=250pt \epsfbox{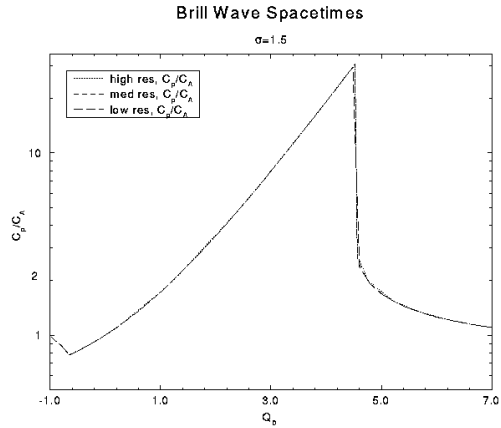}
\caption{
We plot the apparent horizon shape parameter $C_p/C_A$, where $C_p$
 and $C_A$ are the polar and ``areal'' circumferences of the surface,
 for a set of non-rotating distorted black hole initial data sets.
 The shape parameter is plotted against $Q_0$, the amplitude of the
 distorting Brill wave.  All spacetimes shown here have the same Brill
 wave width, $\sigma=1.5$, the width that best characterizes the
 geometry of the Kerr spacetime.  This plot shows the same data as
 computed on three grids with increasingly more zones, which
 illustrates how close we are to the exact solution.  These results
are second order convergent.}
\label{Plot1}
\end{figure}

First we begin by examining the shape parameter, $C_p/C_A$, for a
series of Brill wave plus black hole spacetimes.  These non-rotating
black holes provide a good starting point for our survey.
Fig.~\ref{Plot1} shows $C_p/C_A$ for a set of black holes with
 $\sigma=1.5$, and the Brill wave amplitude, $Q_0$, in the range $-1<Q_0<7$.

The apparent horizon generally becomes prolate if it is distorted with
a positive amplitude Brill wave, and oblate if it is distorted by a
negative amplitude Brill wave~\cite{Bernstein93a}.  However, the shape
of the horizon undergoes two sharp transitions, one at about
$Q_0=-.65$~\cite{Bernstein93a} and the other at about $Q_0 = 4.5$.
Between these two values of $Q_0$ the horizon is on the throat, and
the ratio $C_p/C_A$ grows exponentially.  The further we move
outside this region the further out the apparent horizon moves, and
the more spherical the black hole becomes.

The spacetime behaves as if increasing amounts of energy were being
deposited in the region somewhat away from the horizon as $|Q_0|$ is
increased.  Beyond a certain critical threshold, one would expect a
new horizon would be formed.  Beyond this range, the spacetime is
still becoming more distorted, but the distortion is driving the
apparent horizon further out on the grid and as a result the
distortion is now primarily interior to the apparent horizon.  Thus
the exterior region becomes more spherical as this parameter is
increased.

\begin{figure}
\epsfxsize=250pt \epsfbox{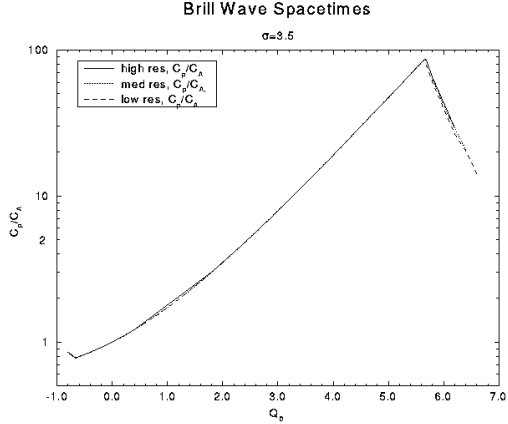}
\caption{
This plot illustrates the basic properties of the apparent horizon shape
parameter, $C_p/C_A$, for the large width Brill wave, $\sigma=3.5$.
The apparent horizon becomes extremely distorted near $Q_0 = 5.65$,
with $C_p/C_A \approx 90$.  This plot shows the same data as computed
on three grids with increasingly more zones.}
\label{Plot2}
\end{figure}

Next we show the effect of changing the width of the Brill wave.  In
Fig.~\ref{Plot2} we show a set of spacetimes like those shown above,
but a larger width Brill wave is used ($\sigma=3.5$).  We see that the
horizon stays on the throat for larger values of $Q_0$, thereby making
the maximum distortion parameter much larger.  This results in a value
of $C_p/C_A$ that is about an order of magnitude larger than the
 $\sigma=1.5$ distortion. The lower transition is still at about
 $Q_0 = -.65$, but now the upper transition has moved to
 $Q_0 =5.65$. The decrease in $C_p/C_A$ as we move away to values of
 $Q_0$ larger than $5.65$ also becomes less sudden.

\begin{figure}
\epsfxsize=250pt \epsfbox{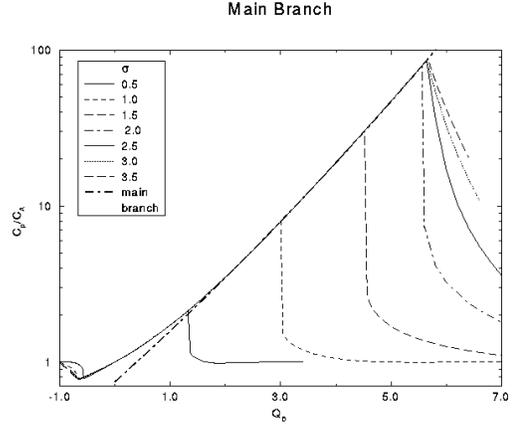}
\caption{
We illustrate the prolateness (large $C_p/C_A$) of the apparent
horizon for spacetimes with Brill wave distortions of varying
amplitude, $Q_0$, and a variety of widths, $\sigma$.  Also plotted is
a fit of the ``main branch,'' a curve on which horizons lie for all
distorted black hole spacetimes considered.  For each family we
consider, there is a transition point, a critical amplitude above
which the horizon jumps off the throat and becomes more spherical.}
\label{Plot3}
\end{figure}

In Fig.~\ref{Plot3} we now show the data from the previous two graphs
along with a number of spacetimes with varying Brill wave widths,
 $\sigma$.  The values of $\sigma$ considered are $0.5$, $1.0$, $1.5$,
 $2.0$, $2.5$, $3.0$ and $3.5$ We will call this group of spacetimes
{\em set 1a}.  We see that the function $C_p/C_A$ of $Q_0$ changes
only slightly as $\sigma$ is varied for a large portion of the plot.
There is a sharp transition that occurs for positive amplitude
distortions which is strongly dependent on the width of the wave.
Beyond this transition point the AH ceases to be on the throat.  There
is another transition point for the negative amplitude Brill wave
spacetimes, and for all but the narrowest of the Brill waves
considered ($\sigma = 0.5$) it lies at about $Q_0 = -0.65$.  The
region between these two transitions constitutes what we will call the
``main branch'' and is relatively independent of $\sigma$.  The ``main
branch'' for the range $1.8<Q_0<5.6$ is approximated to an accuracy of
$1\%$ or less by
\begin{equation}
C_p/C_A ={\rm exp}\left(
-0.287055 + 0.728645\,Q_0 + 0.019728\,{{Q_0}^2}\right),
\end{equation}
and is plotted by a heavy dashed line in Fig.~\ref{Plot3}.  Since the
horizon is simply the throat ($\eta=0$) for the ``main branch''
spacetimes it is not surprising that $\log(C_p/C_A)$ should depend
linearly on $Q_0$ as it is simply reflecting the effect of $Q_0$ upon
the conformal metric.

Also worthy of note is that above $\sigma = 2.0$ the upper transition
point becomes nearly fixed at $Q_0 = 5.65$.  Some small variation
occurs, but not much. Roughly $\sigma=2$ is a width above which the
most prolate spacetime occurs at about $Q_0=5.6$ with a shape
characterized by $C_p/C_A\approx 88$.  Likewise, it appears that
$\sigma=1$ is a width above which the most oblate spacetime occurs at
about $Q_0 = -.65$ with a shape characterized by $C_p/C_A \approx
.77$.

\begin{figure}
\epsfxsize=250pt \epsfbox{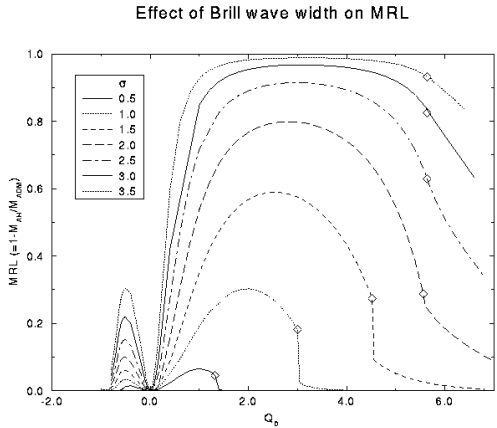}
\caption{
This plot illustrates the maximum radiation loss, or MRL, for
spacetimes with Brill wave distortions of varying amplitude, $Q_0$,
and a variety of widths, $\sigma$.  The diamonds mark the location of
the maximally prolate horizon for each curve}
\label{MRLplot3}
\end{figure}

In Fig.~\ref{MRLplot3} we see the effect of $Q_0$ and $\sigma$ on the
MRL (maximum radiation loss) of the spacetime for {\em set 1a}.  For
reference we have marked the positive $Q_0$ transition points, the
peaks in the plot of $C_p/C_A$ that are seen in Fig.~\ref{Plot3}, with
diamonds in Fig.~\ref{MRLplot3}. The MRL increases, in all cases, as
we increase $|Q_0|$ until it reaches some maximum value.  As we
increase $|Q_0|$ further, we arrive at a transition point.  Along the
path of increasing $|Q_0|$, the energy of the spacetime first
increases primarily because we are adding wave energy outside the
horizon.  As we travel further in this parameter space the mass
increases primarily because energy is being added to the hole itself.
Another feature that can be observed in Fig.~\ref{MRLplot3} is that
the MRL of the transition point begins to rise much more rapidly with
increasing $Q_0$ after the Brill wave width becomes greater than $2$.

In both Fig.~\ref{Plot1} and Fig.~\ref{Plot2} we have plotted the
curves for Brill waves with widths $\sigma=1.5$ and $\sigma=3.5$
respectively at three different grid resolutions:
 $600\times 96$, $300\times 48$, and $150\times 24$.  We note
that in both cases the resolution lines are very close together and
that the convergence exponent, $\xi$, as defined by,
\begin{equation}
\left(\frac{C_p}{C_A}\right)_{num} -
\left(\frac{C_p}{C_A}\right)_{true} \propto
\left( \Delta \eta \right)^\xi+...
\end{equation}
is very nearly $2$ at all points on both graphs.  The larger width
Brill waves were more difficult to calculate, but our code is capable
of calculating these spacetimes accurately.

\subsection{Negative amplitude Brill wave distortions}
In this section we focus on the region of the time-symmetric distorted
black hole spacetime in which the amplitude of the distortion is
negative.  In this regime we find a behavior that is qualitatively
similar to the behavior seen for positive amplitude distortions in
{\em set 1a}, but we found that we needed to use narrower Brill waves
to explore this effect for negative amplitude spacetimes.  If we
consider Fig.~\ref{Plot3} we see that all but the $\sigma=0.5$ curves
are very similar for the negative amplitude portion of the graph.
Because of this we use widths $\sigma=0.3$, $0.4$, $0.5$, ... $0.9$
and we plot this data in Fig.~\ref{obbws}.  We label this data as
{\em set 1b}.

Note in Fig.~\ref{obbws} that the maximum distortion (minimum
 $C_p/C_A$) is increased by widening the Brill wave until a limiting
 value is reached, a behavior similar to that observed for positive
 amplitudes in Fig.~\ref{Plot3}.  There is also a continuation of the
 main branch for the segment of the spacetimes in which the wave is on
 the throat, although we have not computed a fit in this case.
\vbox{
\begin{figure}
\epsfxsize=250pt \epsfbox{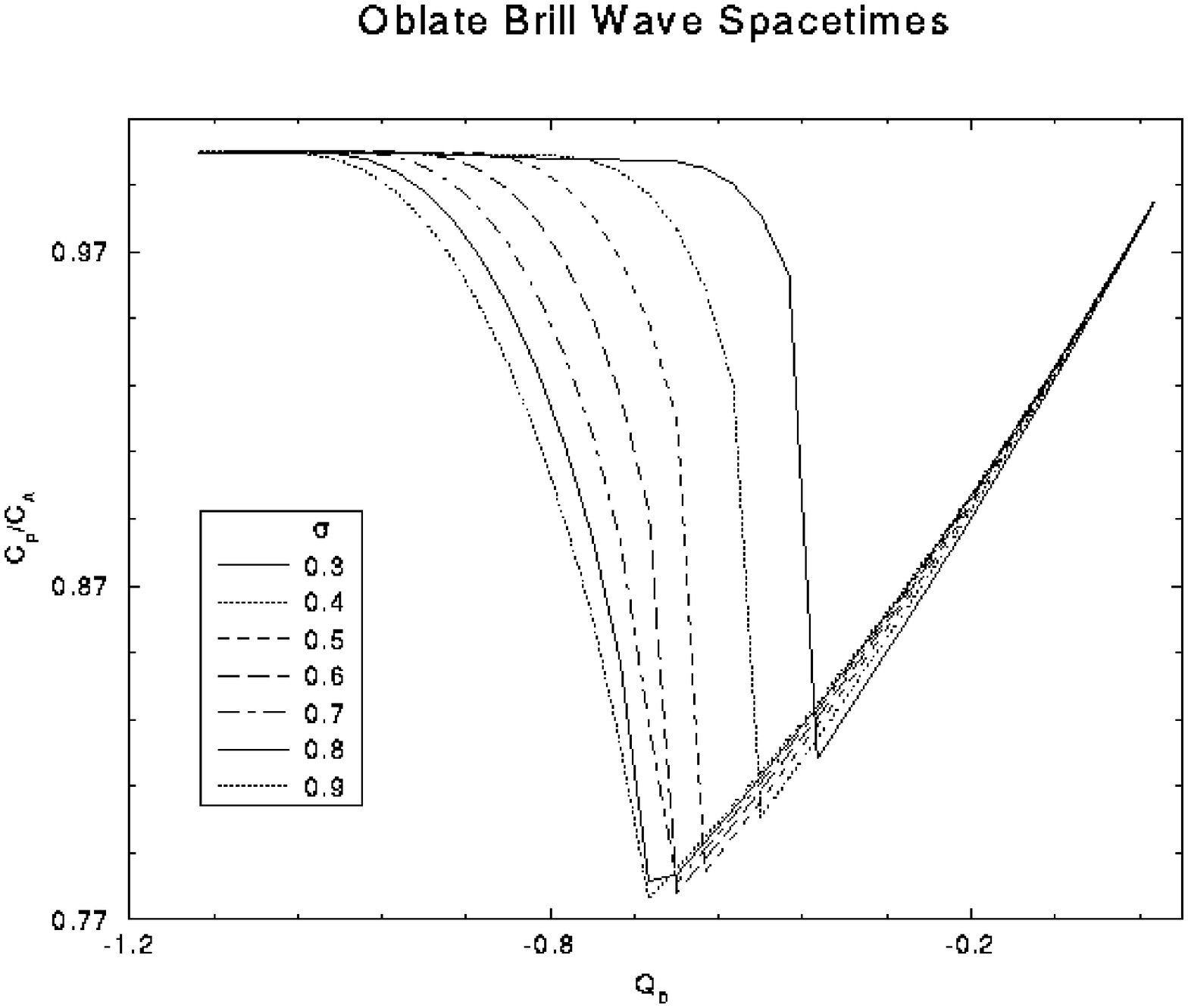}
\caption{
This plot illustrates the behavior of the negative amplitude, narrower
width Brill wave spacetimes.  The behavior here is qualitatively
similar to the larger width positive amplitude spacetimes depicted in
Fig.~4, except that the apparent horizons are all oblate.  As $Q_0$
becomes more negative the apparent horizon jumps off the throat at a
transition point and the spacetime becomes more spherical. }
\label{obbws}
\end{figure}
}

\subsection{Brill wave distortions of Bowen and York rotating black
holes}
\begin{figure}
\epsfxsize=250pt \epsfbox{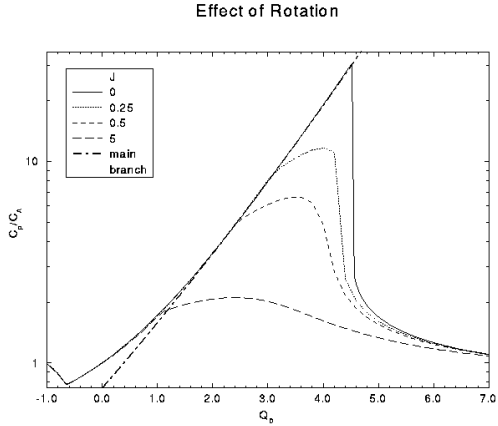}
\caption{
This figure illustrates the effect on the apparent horizon of a Brill
wave distortion on a Bowen and York black hole.  The addition of
angular momentum makes the transition point, the point where the
horizon leaves the throat, occur sooner, although for small values of
 $Q_0$ the value of $C_p/C_A$ is very close to the ``main branch.'' }
\label{rot}
\end{figure}

In this section we discuss spacetimes with $\sigma=1.5$,
 $-1<Q_0<7$ and angular momentum parameter $J=0.25$, $J=0.5$, and
 $J=5$, and we will call this collection {\em set 2a}.  We plot
 $C_p/C_A$ for this set in Fig.~\ref{rot}. One can see from the figure
 that the rotating black holes still lie along the main branch, but
 the transition points occur earlier (for smaller $Q_0$).

\begin{figure}
\epsfxsize=250pt \epsfbox{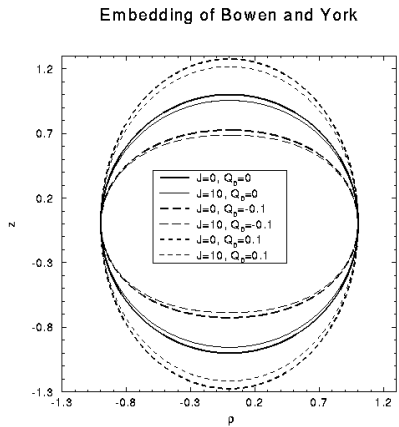}
\caption{
In this plot we see the embeddings of several distorted black hole
spacetimes. The heavy lines are spacetimes distorted by Brill waves: a
prolate spacetime with $Q_0=0.1$, an oblate spacetime with $Q_0=-0.1$,
and a Schwarzschild spacetime with $Q_0=0.0$.  The lighter lines
represent these same spacetimes with angular momentum added.}
\label{Fig:embeddings}
\end{figure}

Adding angular momentum to Brill wave spacetimes reduces the value of
 $C_p/C_A$, thereby making the spacetime more oblate.  For the larger
 amplitude Brill waves this effect is quite marked. In
 Fig.~\ref{Fig:embeddings} we make this point clearer by graphically
 depicting the embeddings of several distorted black hole spacetimes.
 Three Brill wave spacetimes are taken as starting points: $Q_0=-0.1$,
 $0.0$ and $0.1$, and are plotted with thick lines.  Next to each of
 these lines is a thinner one depicting the shape of the horizon after
 setting $J=10$.  This general trend, that angular momentum makes an
 horizon more oblate was a feature that was observed previously by
 Smarr for the Kerr spacetime~\cite{Smarr73b}.

\begin{figure}
\epsfxsize=250pt \epsfbox{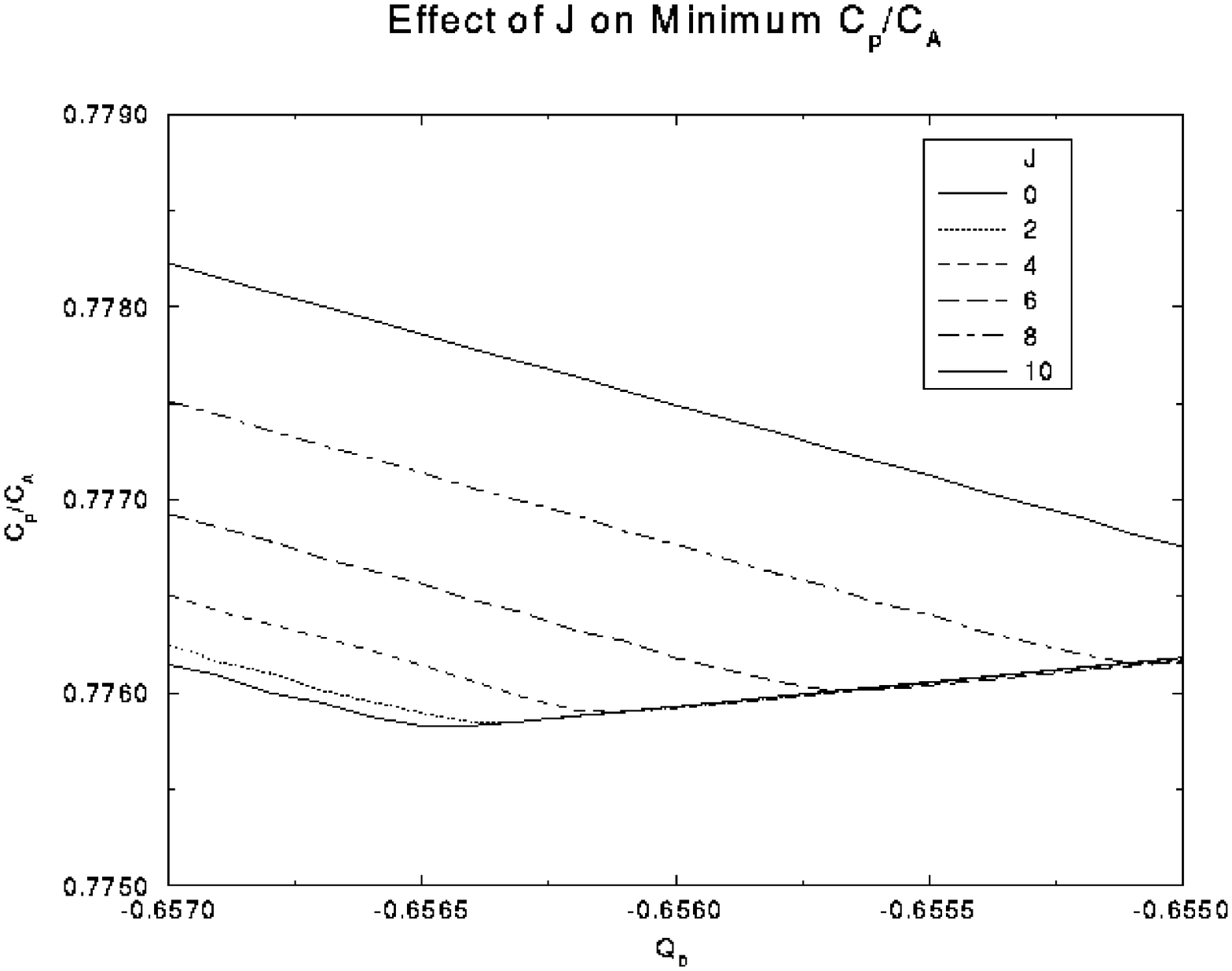}
\caption{
In this plot we see data sets which cover input parameters in the
range $-0.657<Q_0<-0.655$, with $\sigma=1.5$ and $0<J<10$ ({\em set 2b}).
It illustrates that although $J$ usually makes an horizon more oblate,
it can also cause the horizon to leave the throat for sufficiently
negative values of $Q_0$ and thereby make it more spherical rather
than more oblate.}
\label{Fig:Oblate}
\end{figure}

As mentioned above, one generally expects that rotating black holes
will be potentially more oblate than non-rotating spacetimes. However,
there is an exception to this.  In Fig.~\ref{Fig:Oblate} we extend the
above plots to show in more detail the effect of rotation on the
negative amplitude spacetimes.  We see data sets which cover input
parameters in the range $-0.657<Q_0<-0.655$, with $\sigma=1.5$ and
 $0<J<10$ ({\em set 2b}).  Thus, while the addition of angular momentum
to a spacetime usually makes an AH more oblate, this is not always
true.  In fact, the addition of the Bowen and York distortion makes
the minimum $C_p/C_A$ as a function of $Q_0$ {\em larger}.  In other
words, while adding a negative amplitude Brill wave to a given black
hole {\em can} make it more oblate, increasing $J$ also makes the
horizon jump off the throat at a smaller value of $|Q_0|$ so that it
does not achieve the maximum oblateness seen with no rotation.

Although we do not provide a plot, we note that as the parameter $J$
is increased for these spacetimes, the MRL decreases.

\subsection{Brill wave distortions of end-stage ``Cosmic Screws''}
\begin{figure}
\epsfxsize=250pt \epsfbox{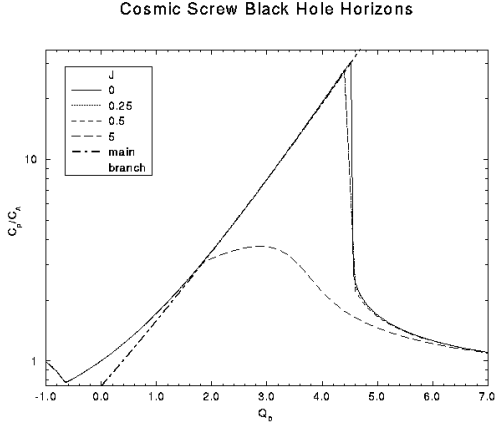}
\caption{
This figure illustrates the effect on the apparent horizon of a Brill
wave distortion on a black hole with a ``Cosmic Screw'' distortion
(counter-rotating above and below the equatorial plane but without net
angular momentum, see text for details) . As in the case of the Bowen
and York black holes, the boundary conditions the Cosmic Screw has the
effect of smoothing out the $C_p/C_A$ curve, although it has a lesser
effect than a Bowen and York distortion with the a distortion
parameter of equal value. }
\label{screw}
\end{figure}

We next examine a collection of ``cosmic screw'' spacetimes, referred
to as {\em set 3}.  The data in {\em set 3} is identical to the data
in {\em set 2a} (distorted Bowen and York black holes) with the
exception that it uses the cosmic screw form of the perturbation.  The
value of $J_{screw}$ clearly creates a much smaller distortion of the
horizon than the Bowen and York angular momentum parameter of the same
value, as is evident in Fig.~\ref{screw}.  Indeed, it is almost
impossible to distinguish the $J_{screw} = 0.5, 0.25$ spacetimes.
This is reasonable since this distortion can be thought of as two
counter-rotating and therefore counter-acting Bowen and York
distortions.

\subsection{Odd-parity distortions of Schwarzschild}
In this section we examine {\em set 4}, which comprises odd-parity
distortions of the form $f_1 = 0$, $f_2 = q^\prime_G$, and
 $f_3=\cos\theta$ (see Eq.~(\ref{theform})), with parameters $\eta_0=0$,
 $0 < Q_0 < 100$, $\sigma = 0.5$, $1.5$, $2.0$, $2.5$, $3.0$, and
 $3.5$.

\begin{figure}
\epsfxsize=250pt \epsfbox{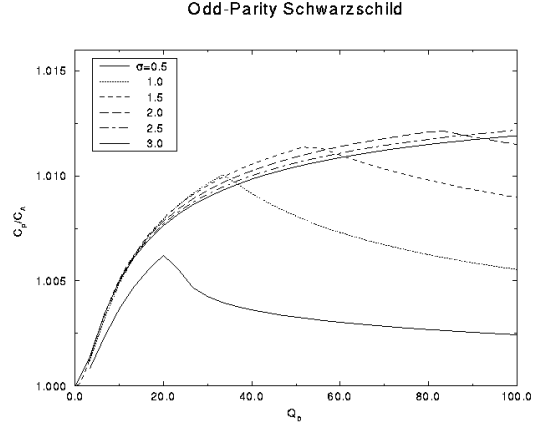}
\caption{
This plot shows the prolateness of the horizon, $C_p/C_A$, for a
odd-parity data sets with a variety of distortion widths, $\sigma$. As
was the case for the Brill wave distorted spacetimes (compare to
Fig. 4 above), the
maximum in the $C_p/C_A$ curve marks the place where the apparent
horizon moves away from the throat. }
\label{Oddcpca}
\end{figure}

This family of spacetimes, depicted in Fig.~\ref{Oddcpca}, displays
the same property of the pure Bowen and York data distortions (in
which $J$, the angular momentum parameter, can be thought of as
representing the amplitude of the extrinsic curvature distortion),
namely that there is no limit to the magnitude of the parameter, $J$,
that is possible, either theoretically or in any numerical test done
to date.  We do not feature the negative amplitude distortions of
these sets because the relevant terms are squared in the Hamiltonian
constraint, making their overall sign unimportant.

Unlike the Bowen and York spacetime, for which the horizon remains on
the throat for very large $J$, the horizon moves away from the throat
in the odd-parity distortion of Schwarzschild for sufficiently large
$Q_0$.  In this respect these data sets are similar to {\em set 1a}.
They also resemble {\em set 1a} in that larger width distortions make
it possible to achieve more distorted horizons.  {\em Set 4} is unlike
{\em set 1a} in that the distortion never becomes even remotely as
large (note that this distortion is in the extrinsic curvature and
affects the 3-metric only through the conformal factor, it is not the
standard Brill wave distortion).

Fig.~\ref{OddMRL} shows the behavior of the MRL function for {\em set
5}.  As in {\em set 1a} this data set has its peak MRL somewhat before
the horizon departs from the throat.  It is otherwise qualitatively
similar to the MRL curves depicted in Fig.~\ref{MRLplot3}.

\vspace{-0.1in}
\vbox{
\begin{figure}
\epsfxsize=250pt \epsfbox{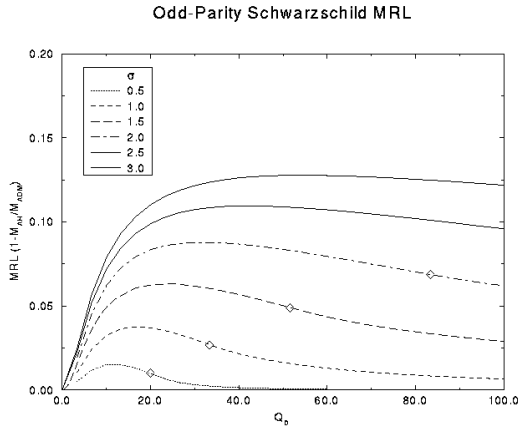}
\caption{
This plot shows the MRL for a variety of odd-parity data sets.
Qualitatively it has the same features observed in Fig.~5, except that
the magnitude of the MRL is much smaller. As in Fig.~5, diamonds mark
the amplitudes at which a given black hole spacetime has its maximum
value of $C_p/C_A$.}
\label{OddMRL}
\end{figure}
}
\subsection{Hoop Conjecture}
An important question that we wish to discuss is ``How distorted can a
black hole be?''  The hoop conjecture~\cite{Thorne72a,MisnerHoopRef}
states that an event horizon will form if and only if the matter is
sufficiently compact in all directions.  This conjecture would seem to
suggest that the black hole horizon should also be sufficiently
compact in all directions. Although interesting work has been done on
the hoop conjecture for black hole spacetimes (see,
e.g.~\cite{Flannagan91,Flannagan92}) much remains to be done.

As we have seen above, black hole {\em apparent} horizons can be
extremely distorted, nonspherical objects as measured by the shape
parameter $C_p/C_A$.  We have shown cases where the AH is extremely
prolate ($C_p/C_A \approx 88$) or oblate ($C_p/C_A \approx 0.77$).
However, these results are for the apparent horizon, and not the {\em
event} horizon.  A careful study of distorted event horizons, and
implications on the hoop conjecture will be presented elsewhere.  In
this section we restrict our attention to the apparent horizons we
have found, and discuss whether there is a limit to how distorted they
can be.

The fact that there are limits on the amplitude parameter ($Q_0$) is
somewhat suggestive of a limit on the degree of distortion possible
for the apparent/event horizon of the black holes considered here
(since the amplitude parameter controls the degree of distortion of
the black hole).  Indeed, it allows one to demonstrate the limit
numerically for a Brill wave of a given shape (as given by $\sigma$,
$\eta_0$, and $n$).

However, there may also be a good reason to suppose that there is no
limit on how large or small the quantity $C_p/C_A$ may become for an
asymptotically flat spacetime.  We might be able to engineer
spacetimes with two gravity wave distortions: one widely extended, the
other more comparable to the size of the hole. In effect, the larger
distortion can be used to ``fool'' the black hole into thinking that
the spacetime is not asymptotically flat. Thus, it may be that we can
find spacetimes with arbitrarily large $C_p/C_A$ if we combine
distortions on a variety of scales.  This possibility is not
investigated here.

\subsection{Maximum Prolateness}
It is difficult to pin down a maximum prolateness for all the types of
spacetimes considered here.  It seems reasonable, however, to simply
focus on Brill wave spacetimes since the odd-parity and momentum
distortions usually make the spacetime more oblate or have small
effect.  For any given value of $\sigma$ it is a straightforward
matter to obtain the value of $Q_0$ which maximizes $C_p/C_A$ because
we can search the entire range of possible values for $Q_0$ (which we
know from section~\ref{bwlimits} must have a value less than $10.0$).

In all these data sets, the horizon remains on the throat until some
critical value of $Q_0$ is reached and then it departs and moves
outward in coordinate space.  Until the horizon leaves the throat,
 $C_p/C_A$ continues to grow, but as soon as it departs this ratio
begins to decrease.  This downward trend in $C_p/C_A$ continues as
 $Q_0$ is increased until a value of $Q_0$ is reached which is
inconsistent with asymptotic flatness.  The most distorted spacetime
occurs when the largest value of $\sigma$ is used, although the growth
is very slow after the value of ${\rm exp}(\sigma)$ has increased
to about half the fundamental $\ell=2$ wavelength of the black hole.

While the information we obtained does not provide us with a definite
answer about the maximum possible prolateness of the apparent horizon
in these spacetimes, it suggests that there is some limiting
value of $C_p/C_A$.  The largest value we have measured is
 $C_p/C_A=88$.  However, it is possible that despite
the observed trends the quantity $C_p/C_A$ may be made to grow without
limit for some untried Brill wave shape or combination of shapes.

\subsection{Maximum Oblateness}
How oblate can a black hole be?  To get a feel for the meaning of the
ratio $C_p/C_A$ we note that for an infinitely thin disk in Euclidean
space we have $C_p/C_A = 0.900$ and for an extremal Kerr black hole
($a/m=1$) we have a smaller value, $C_p/C_A = 0.860$.  Thus, although
these ratios are quite close to unity, they actually represent
extremely distorted surfaces.  For a very wide range of Brill wave
parameters we have seen a robust lower bound for $C_p/C_A \approx
0.77$, with $Q_0 = -0.65$, as discussed above.  This seems to follow
the trend discussed above for the prolate black hole spacetimes.

\subsection{Penrose Inequality}
\label{penrose}
In 1973 Penrose proposed a criterion for a spacetime which, if
violated, would indicate Cosmic Censorship would be
violated~\cite{Penrose73}.  The criteria is simply that the
irreducible apparent horizon mass must be less than or equal to the
ADM mass.

Jang and Wald proved~\cite{Jang77} the inequality for spacetimes with
a single AH of topology $S^2$ which also satisfies a certain
mathematical condition. (It essentially states that it is possible
to foliate the region of the spacetime exterior to the apparent
horizon with a sequence of nested 2-surfaces, which are
asymptotically spheres at large radii, using a ``lapse'' whose value is
equal to the inverse of the extrinsic curvature of those 2-surfaces)
We have not shown that this condition is actually satisfied for the
wide class of black hole spacetimes considered here, but we have seen
that this inequality is preserved in all spacetimes considered in this
paper. Bernstein also sought for and failed to find a violation of the
Penrose inequality~\cite{Bernstein93a,Bernstein94a} using the Brill
wave distortions of the non-rotating black hole, but did not consider
the extreme distortions used in this paper.
A violation of the Penrose inequality would result in a negative value
for the MRL, and we can see that this does not occur if we review
Figs.~\ref{MRLplot3} and \ref{OddMRL}.

\section{Summary and future work}
\label{summary}
In this paper we have described the construction of new initial
data sets for distorted rotating black holes in detail.  These data
sets correspond to a large family of black hole spacetimes,
generalizing the Kerr\cite{Kerr63}, Bowen and York~\cite{Bowen80},
distorted Schwarzschild black holes~\cite{Bernstein94a}, and should
provide a rich testing ground for evolutions of highly dynamic,
rotating and nonrotating black holes.

We have analyzed many of the properties of extreme parameter choices
these data sets.  We have, for example, located the analytic upper and
lower limits to the possible distortion parameters for these
spacetimes, located the transition points beyond which multiple
trapped surfaces exist, and identified spacetimes with the most
oblate/prolate apparent horizons in our family of data sets.
Furthermore, we investigated the effect of rotation and cosmic-screw
type distortions on these data sets.  This investigation should help
provide a roadmap to interesting, distorted black hole spacetimes for
evolution in both 2D and 3D.

In the near future we intend to apply the horizon finding code
described in Ref.~\cite{Libson94a} to some of the above data sets, to
determine to what extent the behavior of the event horizons is in
agreement with the behavior of the more extremely distorted apparent
horizons.  Do spacetimes with strongly distorted apparent horizons
have strongly distorted event horizons?  This research provides us
with important data for this investigation.  In particular, $\sigma=2$
configurations with amplitudes $Q_0>3.5$ seem likely to yield the most
distorted spacetimes.  We also plan to evolve the cosmic screw
spacetimes and study them in more detail.

\section{Acknowledgements}
We would like to thank Pete Anninos, David Bernstein, Karen Camarda, 
Greg Cook, Paul Walker, Larry Smarr, and especially Wai-Mo Suen for 
many helpful suggestions throughout the course of this work. This 
work was supported by NCSA, NSF grants PHY94-07882 and 
ASC/PHY93-18152 (ARPA supplemented). The calculations were performed 
at NCSA on the SGI Power Challenge and at the Pittsburgh 
Supercomputing Center on the Cray C-90. Some calculations were done 
using MathTensor and Mathematica.

\end{document}